\pdfoutput=1 
\documentclass[notoc]{JINST}
\let\ifpdf\relax
\usepackage{xspace}
\usepackage{url}
\usepackage{paralist}
\newcommand{\apenetp}{APEnet+\xspace}
\newcommand{\apelink}{APElink\xspace}

\newcommand{\nanet}{NaNet\xspace}
\newcommand{\nanetone}{NaNet-1\xspace}
\newcommand{\nanetten}{NaNet-10\xspace}
\newcommand{\nanetcube}{NaNet$^3$\xspace}
\newcommand{\gbe}{GbE\xspace}

\newcommand{\realtime}{real-time\xspace}
\newcommand{\lowlatency}{low-latency\xspace}
\newcommand{\nvidia}{NVIDIA\xspace}
\newcommand{\pcie}{PCIe\xspace}
\newcommand{\nios}{Nios~II\xspace}

\newcommand{\eg}{\textit{e.g.}\xspace}
\newcommand{\tengbe}{10-GbE\xspace}

\title{\nanet: a flexible and configurable \lowlatency NIC for
  \realtime trigger systems based on GPUs.}

\author{R. Ammendola$^a$, A. Biagioni$^b$, O. Frezza$^b$,
  G. Lamanna$^c$$^d$, A. Lonardo$^b$\thanks{Corresponding author.}~, F. Lo
  Cicero$^b$, P.~S.~Paolucci$^b$, F. Pantaleo$^c$, D. Rossetti$^b$,
  F. Simula$^b$, M. Sozzi$^c$$^d$, L. Tosoratto$^b$, P. Vicini$^b$\\
  \llap{$^a$}INFN Sezione di Roma Tor Vergata,\\
  Via della Ricerca Scientifica, 1 - 00133 Roma, Italy\\
  \llap{$^b$}INFN Sezione di Roma\\
  P.le Aldo Moro, 2 - 00185 Roma, Italy\\
  \llap{$^c$}INFN Sezione di Pisa\\
  Via F. Buonarroti 2 - 56127 Pisa, Italy\\
  \llap{$^d$}CERN\\
  CH-1211 Geneva 23, Switzerland\\
  E-mail: \email{alessandro.lonardo@roma1.infn.it}}

\abstract{\nanet is an \mbox{FPGA-based} \pcie X8 Gen2 NIC supporting
  1/10~\gbe links and the custom 34~Gbps \apelink channel. The design
  has \textit{GPUDirect RDMA} capabilities and features a network
  stack protocol offloading module, making it suitable for building
  \lowlatency, \realtime GPU-based computing systems. We provide a
  detailed description of the \nanet hardware modular
  architecture. Benchmarks for latency and bandwidth for \gbe and
  \apelink channels are presented, followed by a performance analysis 
  on the case study of the \mbox{GPU-based} low level trigger for the RICH detector in the NA62 CERN experiment,
  using either the \nanet \gbe and \apelink channels.
  Finally, we give an outline of project future activities.}


\begin{document}

\section{Introduction}
\label{sec:intro}
Thanks to their relevant computing power and favorable ratios in
price/performance and power consumption/performance, GPUs
architectures such as \nvidia Fermi and Kepler are gaining popularity
in the HEP experiments community.
Their usage in high level trigger systems, leveraging on their
computing power to reduce the numerosity of computing farm nodes, is
currently under study with encouraging
results~\cite{Clark:2010:GPUsAtlas, Rohr:2012:GPUsAlice, Halyo:2013:GPUsCMS}.
For the same reasons, low level triggers could also benefit from GPUs
adoption; the main issue to be taken into account in this context is
the strict \realtime requisite typical of such systems.
%

Low level triggers are designed to perform very rough selection based
on a \mbox{sub-set} of the available information, in a  pipelined
structure housed in custom electronics, in order to bring to
a manageable level the high data rate that would otherwise reach the
software stages behind them.
Due to small buffers size in the \mbox{read-out} electronics, such
systems typically require very low latency; however, thanks to fast
and cheap DDR memories available nowadays, this requirement will be
abandoned in the near future.
%
%
On the other hand, GPUs provide so great a computing power that taking
complex decisions with speeds matching significant data rates is
feasible; this would mean more accurate selection and more stringent
trigger conditions, providing purity and efficiency such as those from
commodity PCs without forfeiting the \realtime constraint. 
At the same time GPUs would represent a great step forward in terms of
reprogrammability when compared to custom electronics.
%
%
%
%
GPUs \realtime performances need careful assessment to match the
requirements of the lowest trigger levels, the main issue being the
network transfer from the custom readout (RO) electronics to the
server hosting the GPU on the \pcie bus.
%
%
Another caveat of GPU architectures is the need for saturation of
computing cores, which requires a significant number of events and a
buffering stage; both factors weigh on trigger answer latency.
Latency stability is another feature that must be carefully considered
for \realtime applications since computing on GPUs is mostly
deterministic as soon as data has landed onto the internal memories
but, wholly considering the low level trigger, latency fluctuations
stem from transit from the RO system through network interface card
(NIC) and \pcie bus.

Our approach to this problem is twofold: first, we designed a NIC able
to inject RO data directly from the links into \nvidia Fermi- and
\mbox{Kepler-class} GPUs memories without any intermediate buffering
or CPU operation --- GPUDirect RDMA is the commercial name of the
feature; second, we implemented a dedicated engine in the NIC to
offload the CPU from network stack protocol management duties.
In this way, transfer latency and its fluctuations are reduced and
possible OS jitter effects avoided.
These two features stand in the \nanet \mbox{FPGA-based} NIC: the
first was inherited from development of our \mbox{HPC-dedicated} 3D
NIC, \apenetp~\cite{ammendola2012apenet+}; the second comes from
adapting and integrating an open core by the FPGA
vendor\footnote{Documentation is here:
  \url{http://www.alterawiki.com/wiki/Nios_II_UDP_Offload_Example}}.

\nanet is flexible, supporting 4 different link technologies, namely a
custom 1~Gbps optical serial link, \gbe
(\mbox{1000BASE-T}/\mbox{1000BASE-X}), \tengbe (\mbox{IEEE 802.3aq})
and the \apelink channel --- 4 bonded PCML lanes over QSFP+ cables
capable of 34~Gbps raw data bandwidth~\cite{APEnetTwepp:2013}; \nanet
logic can be effectively tailored to different usage scenarios as any
\mbox{FPGA-based} design by adding dedicated custom logic blocks, \eg
to compress or reshuffle the data stream.

\nanet is currently being used in a pilot project within the CERN NA62
experiment aiming at investigating GPUs usage in the central Level 0
trigger processor (L0TP)~\cite{Lamanna:2009:GPUsNA62short}.

In the following we provide a detailed description of the \nanet
hardware modular architecture and a performance analysis
for a case study on the \mbox{GPU-based} Level 0 trigger of the NA62
RICH detector using either the \nanet \gbe and \apelink channels.

Results of this study motivated current development of \nanet design
aimed at including 10~\gbe link support; preliminary results and
additional FPGA resources requirements are shown.

Finally, we report an outline of future project developments.

 
%
\begin{figure}[b]
  \begin{minipage}[t]{.49\textwidth}
    \centering
    \includegraphics[width=\textwidth]{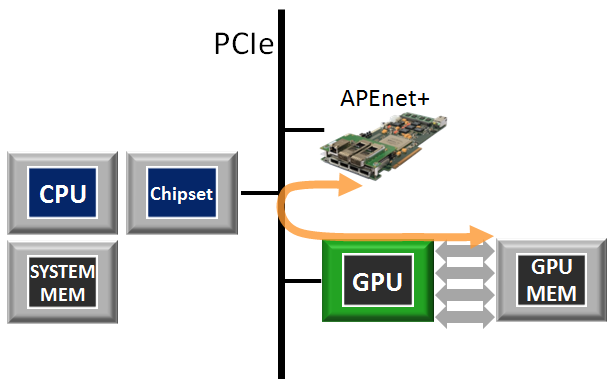}
    \caption {\apenetp/GPUDirect RDMA.}
    \label{fig:apenetflow}
  \end{minipage}
  \quad
  \begin{minipage}[t]{.49\textwidth}
    \centering 
    \includegraphics[trim=60mm 25mm 60mm 40mm, clip,width=.8\textwidth]{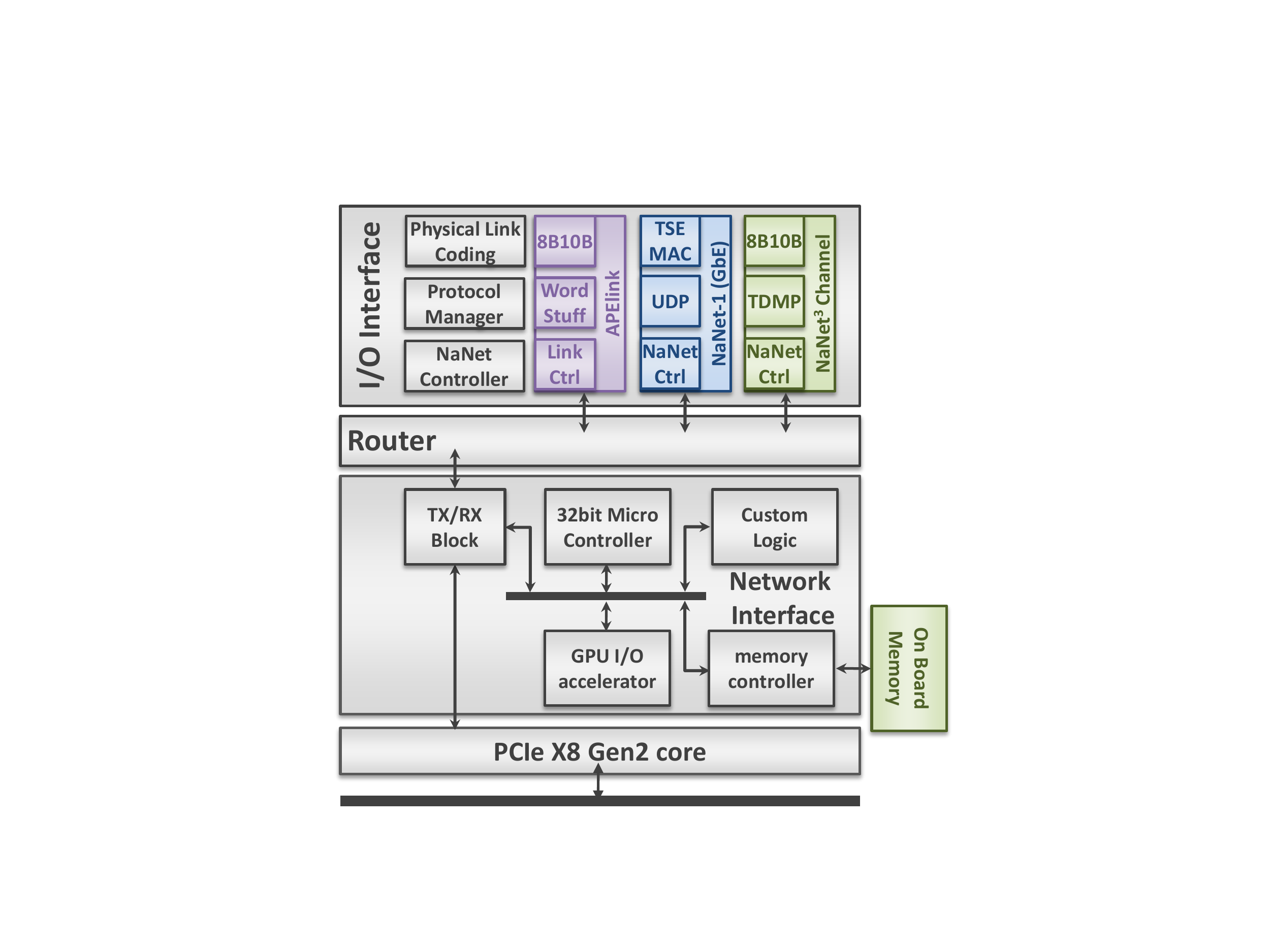}
    \caption{\nanet architecture schematic.}
    \label{fig:NaNet}
  \end{minipage}
\end{figure}

\section{\nanet}
\label{sec:nanet}
\nanet is a modular design of a \lowlatency NIC dedicated to \realtime
GPU-based systems and supporting a number of different physical links;
its design baseline comes from the \apenetp \pcie Gen 2 x8 3D
NIC.
The \textit{Distributed Network Processor} (DNP) is the \apenetp core
logic, acting as an \mbox{off-loading} engine for the computing node
in performing \mbox{inter-node} communications~\cite{DNP2012short}.
The DNP provides hardware support for the Remote Direct Memory Access
(RDMA) protocol guaranteeing \lowlatency data transfers.
Moreover, \apenetp is also able to directly access the Fermi-
and \mbox{Kepler-class} \nvidia GPUs memory (provided that both devices
share the same upstream \pcie root complex) leveraging upon their
\mbox{peer-to-peer capabilites}.
This is a \mbox{first-of-its-kind} feature for a non-\nvidia device
(GPUDirect RDMA being its commercial name), allowing unstaged
\mbox{off-board} \mbox{GPU-to-GPU} transfers with unprecedented low
latency~\cite{ammendola:2013:GPUshort}.
An overview of the typical \apenetp data flow is in
figure~\ref{fig:apenetflow}: inward and outward traffic over the
34~Gbps \apelink channel is directly routed to and from GPU internal
memory.

\nanet design inherits GPUDirect RDMA capabilities from \apenetp,
extends it with support for standard network links --- namely \gbe and
10~\gbe --- and adds to the logic a network stack protocol management
offloading engine, to avoid possible OS jitter effects and reduce
latency even more.
\nanet design supports a configurable number and kind of I/O channels;
incoming data streams are processed by a Physical Link Coding block
feeding the Data Protocol Manager that in turn extracts the payload
data.
These payload data are encapsulated by the \nanet Controller in the
\apenetp data packet protocol and sent to the \apenetp Network
Interface, taking care of their delivery to the destination memory.
A Custom Logic block joins in by performing any data manipulation
needed by the specific application context (see
figure~\ref{fig:NaNet}).
%
In the following, we focus on the characterization of the \nanetone
design configuration, then we describe current developments for one
supporting 10~\gbe interface, \nanetten; finally, we present a sketch
of the \nanetcube design, its main feature being its deterministic
latency links.
 
\subsection{\nanetone architectural overview}
\label{sec:nanetone}
\begin{figure}[b]
  \begin{minipage}[t]{.49\textwidth}
    \centering
    \includegraphics[width=\textwidth]{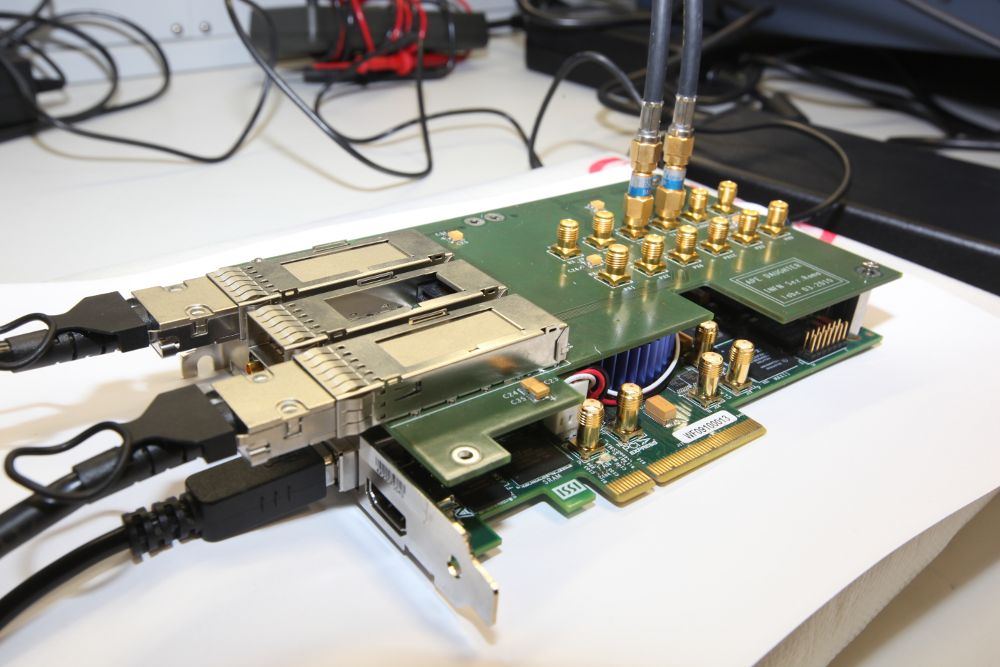}
    \caption{\nanetone on Altera Stratix IV dev. board with custom
      mezzanine card + 3 \apelink channels.}
    \label{fig:board}
  \end{minipage}
  \quad
  \begin{minipage}[t]{.49\textwidth}
    \centering 
    \includegraphics[width=\textwidth]{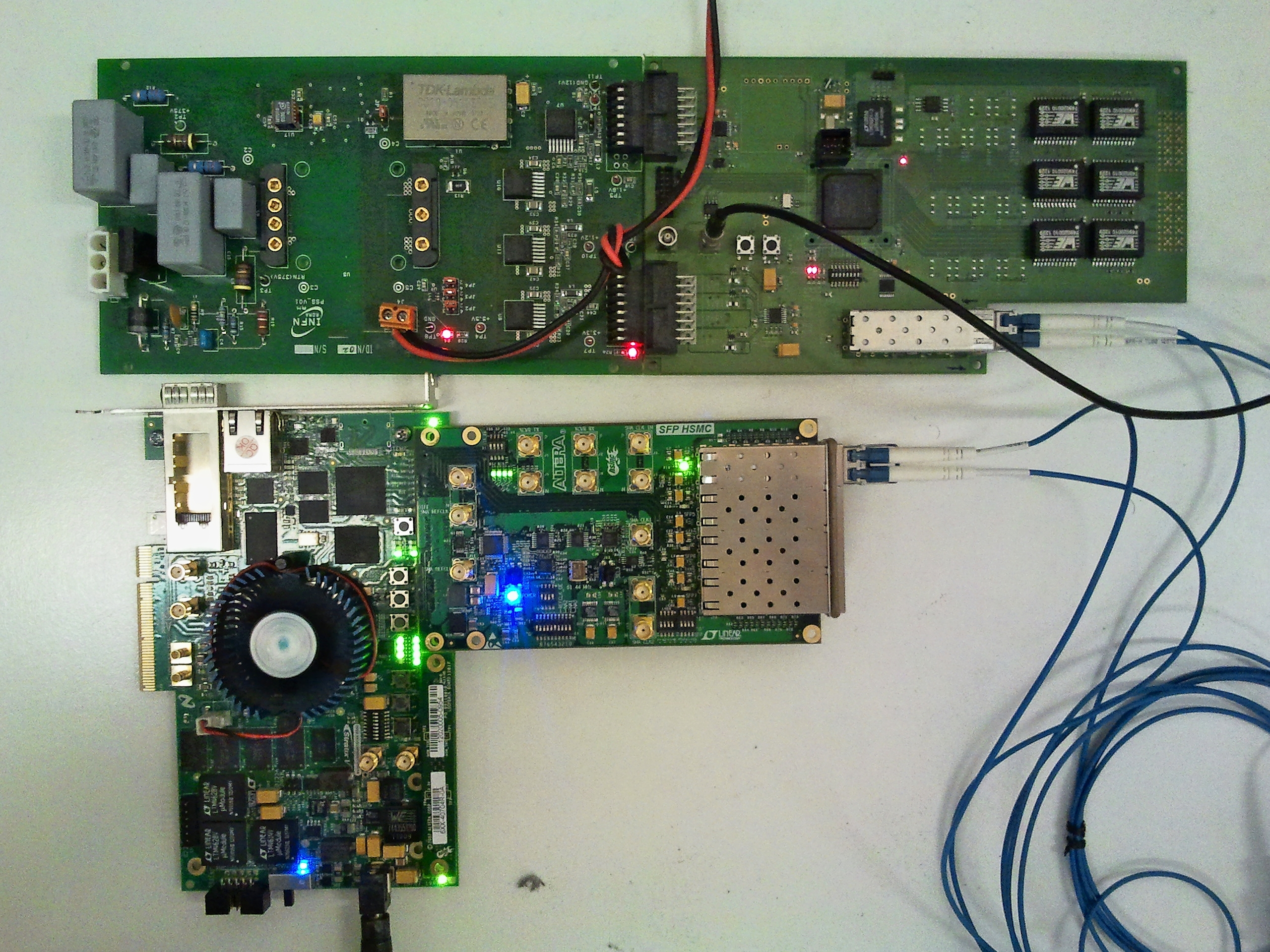}
    \caption{\nanetcube testbed: board is connected to offshore RO
      system via optical cable.}
    \label{fig:nanetcube}
  \end{minipage}
\end{figure}

\nanetone is a \pcie Gen 2 x8 NIC featuring GPUDirect RDMA over 1 \gbe
and optionally 3 \apelink channels.
The \nanetone board employs the Altera Stratix IV EP4SGX230KF40C2 FPGA
(see figure~\ref{fig:board}); a custom mezzanine was designed to be
optionally mounted on top of the Altera board.
The mezzanine mounts 3 QSFP+ connectors, thus making \nanet able to
manage 3 \mbox{bi-directional} \apelink channels with switching
capabilities up to 34~Gbps.
\apelink adopts a proprietary data transmission word stuffing
protocol; this is pulled for free into \nanetone.

For what concerns the implementation of the \gbe transmission system
we follow the general I/O interface architecture description of
figure~\ref{fig:NaNet}.

We exploit the Altera Triple Speed Ethernet Megacore (TSE MAC) as
Physical Link Coding, providing complete 10/100/1000~Mbps Ethernet IP
modules.
The design employs SGMII standard interface to connect the MAC to the
PHY including Management Data I/O (MDIO); the MAC is a single module
in FIFO mode for both the receive and the transmit sides ($2048x32$
bits).

The data protocol manager tasks are carried out by the \textit{UDP
  Offloader} dealing with UDP packets payload extraction and providing
a \mbox{32-bit} wide channel achieving 6.4~Gbps (6 times greater than
the standard \gbe requirements).
The UDP Offloader component collects data coming from the Avalon
Streaming Interface of the Altera Triple Speed Ethernet Megacore and
redirects UDP packets into a hardware processing data path.
In this way, the FPGA \mbox{on-board} $\mu$controller (\nios) is
totally discharged from UDP packet traffic management.

The I/O interface data flow control logic is managed by the
\textit{\nanet Controller}, a hardware component able to encapsulate
data packets in the \apenetp protocol formed by a
\textit{header}, a \textit{footer} (\mbox{128-bit} word) and a
\textit{payload} of maximum size equal to 4096~bytes.
\nanet Controller implements an \mbox{Avalon-ST} Sink Interface
collecting the \gbe data flow from the UDP offloader, parallelizing
incoming \mbox{32-bit} data words into \mbox{128-bit} \apenetp data
ones.

Data coming from the I/O interface are managed by the \textit{Router}
component; it supports a configurable number of channels, acting as a
multiplexer for a customizable number of ports.

Finally, the \textit{Network Interface} comprises the \pcie X8 Gen2
link to the host system for a maximum data rate of \mbox{4+4}~GB/s,
the packet injection processing logic, the \textit{RX block} and
\textit{GPU I/O accelerator} providing hardware support for the RDMA  
protocol for CPU and GPU, managed by the \nios $\mu$controller operating at 200~MHz.
On table~\ref{tab:logic} we show a recap of the used FPGA logic
resources as measured by the synthesis software.

\begin{table}[b]
  \caption{An overview of \nanet resource consumption.}
  \label{tab:logic}  
  \smallskip
  \centering
  \begin{tabular}{|c|cccc|}
    \hline
    \textbf{Project}  & \textbf{Board}  & \textbf{Comb. ALUT} & \textbf{Register} & \textbf{Memory [MB]}  \\
    \hline                           
    \nanetone         & EP4SGX230KF40C2 & 54635 (30\%)        & 54415 (30\%)      &   1.00 (55\%)         \\
    \nanetten         & EP4SGX230KF40C2 & 51325 (28\%)        & 49378 (27\%)      &   0.97 (53\%)         \\
    \hline
    \hline
    \multicolumn{2}{|c|}{\textbf{Logic Block}}   & \textbf{Comb. ALUT}   & \textbf{Register}       & \textbf{Memory [MB]}      \\
    \hline
    \multicolumn{2}{|l|}{PCIe}                   & 7268         & 8042      & 0.001                 \\
    \multicolumn{2}{|l|}{Network Interface}      & 37130        & 38226     & 0.877                 \\
    \multicolumn{2}{|l|}{Router}                 & 4652         & 3855      & ---                   \\
    \multicolumn{2}{|l|}{\apelink I/O}           & 13820        & 13700     & 0.131                 \\
    \multicolumn{2}{|l|}{\gbe I/O}               &  543         & 575       & 0.066                 \\    
    \multicolumn{2}{|l|}{\tengbe I/O}            & 8395         & 7499      & 0.005                 \\    
    \hline
  \end{tabular}
\end{table}
\subsection{Software Stack}
\label{sec:sw}
The \nanetone software stack runs partly on the x86 host and partly on
the \nios \mbox{FPGA-embedded} \mbox{$\mu$controller}.
On the host side a GNU/Linux kernel driver controls the device and an
application level library provides an API to: open/close the \nanetone
device; inject commands to register and \mbox{de-register} circular lists
of persistent receiving buffers (CLOPs) in GPU and/or host memory,
necessary to allocate, pin and return the virtual address of these
buffers to the application; manage events generated by the device when
receiving packets on the registered buffers in order to promptly
invoke the GPU kernel that processes the data just received.
On the $\mu$controller, a single process C program configures the
device, computes the destination virtual address inside the CLOP for
incoming packets payload and performs the virtual to physical memory
address translation necessary to initiate the \pcie DMA transaction
towards the destination buffer.

\subsection{\nanetone enhancements and roadmap to \nanetten}
\label{sec:nanetten}
As described in section~\ref{sec:sw}, the \nanet GPU memory addressing
is managed by the \nios firmware.
Implementing new features with a $\mu$controller is a fast and
efficient strategy during debugging phase but the \nios introduces a
considerable latency in performing the basic RDMA tasks: buffer search
and translation of virtual addresses to physical ones.
Moreover, it is responsible of jitter effects on the hardware latency path~\cite{CHEP2013_arXiv:NANETshort}.
Thus, two major improvements are currently under development for
\nanetone: a Translation Lookaside Buffer (TLB), an associative cache
where a limited amount of entries can be stored in order to perform
memory management tasks, taking only $\sim200$~ns and a hardware module for virtual address generation for GPU memory
management.

The expected request of increased data rates and considerations of
\mbox{future-proofing} for the \nanet IP pushed the design of a board
supporting the more advanced \tengbe industrial standard: \nanetten.
Since Altera Stratix IV development board is not natively equipped
with a \tengbe interface, an additional board from Terasic (Dual XAUI
To SFP+ HSMC) is employed; it mounts a Broadcomm BCM8727
\mbox{dual-channel} \tengbe \mbox{SFI-to-XAUI} transceiver and
provides 2 full duplex \tengbe channels with a XAUI backend interface.
This mezzanine card is plugged into the HSMC connector of the Altera
board.
At the moment, this makes the \tengbe mutually exclusive with the
custom mezzanine providing the \apelink channels.

The final configuration foresees design migration towards Stratix V
FPGA, to exploit enhanced Altera transceivers with switching
capabilities up to 12.5~Gbps and a \mbox{Gen3-compliant} \pcie bus
able to sustain $8+8$~GB/s.
\subsection{\nanetcube four way deterministic latency 1 Gbps optical link NIC}
\label{sec:nanetcube}
To be complete, an overview of the \nanet board family must mention
the undergoing development of the \nanetcube board for the KM3 HEP
experiment~\cite{KM3TechRep:2010}.
In KM3 the board is tasked with delivering global clock and
synchronization signals to the underwater electronic system and
receiving photomultipliers data via optical cables.
The design employs Altera Deterministic Latency Transceivers with an
8B10B encoding scheme as Physical Link Coding and Time Division
MultiPlexing (TDMP) data transmission protocol.
Current implementation is being developed on the Altera Stratix V
development board with a Terasic \mbox{SFP-HSMC} daughtercard plugged
on top and sporting 4 \mbox{transceiver-based} SFP ports (see
figure~\ref{fig:nanetcube}).

\section{\nanetone performances}
\label{sec:perf}

\nanetone performances were assessed on a Supermicro SuperServer
\mbox{6016GT-TF}.
The setup comprised a \mbox{X8DTG-DF} (Tylersburg chipset --- Intel
5520) dual socket motherboard, 2 Intel 82576 \gbe ports and \nvidia
M2070 GPU; sockets were populated with Intel Xeon X5570 @2.93~GHz.

Measurements were conducted using one of the host \gbe ports to send
UDP packets according to the NA62 RICH RO data protocol to the
\nanetone \gbe interface: using the x86 Time Stamp Counter (TSC) register as a common
time reference, it was possible in a single process test application
to measure latency as time difference between when a received buffer
is signalled to the application and the moment before the first UDP
packet of a bunch (needed to fill the receive buffer) is sent through
the host \gbe port.
Similarly, we closed in a loopback configuration 2 of the 3 available
\apelink ports and performed the same measurement.
Note that in the aforedescribed measurement setup (``system
loopback''), the latency of the send process is also taken into
account.
\begin{figure}[t]
  \begin{minipage}[t]{.49\textwidth}
    \includegraphics[trim=20mm 20mm 10mm 15mm,clip,width=\textwidth]{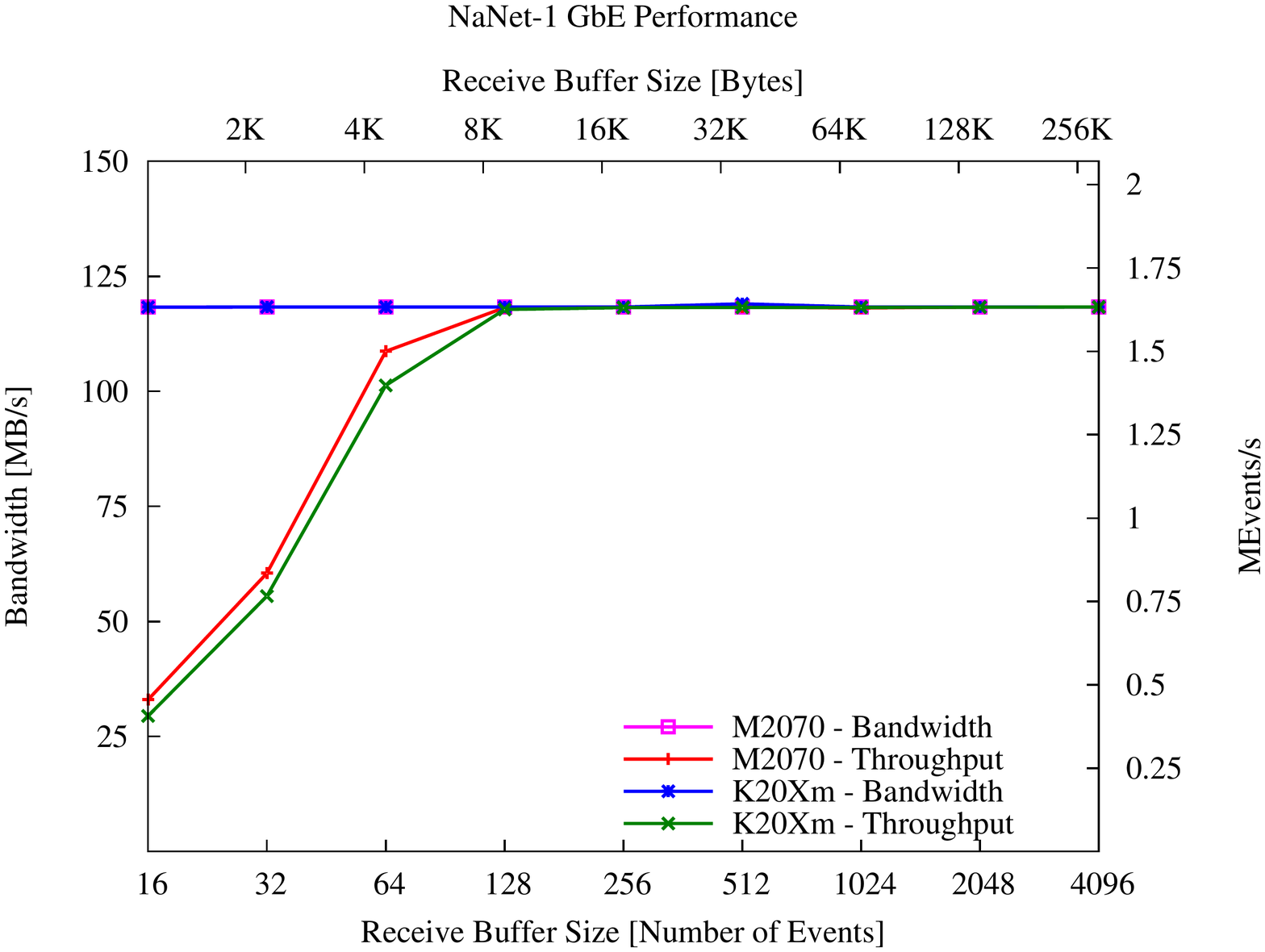}
    \caption {\nanetone \gbe I/O bandwidth and \mbox{\gbe-fed} RICH L0TP
      throughput.}
    \label{fig:banda_throughput_nanet}
  \end{minipage}
  \quad
  \begin{minipage}[t]{.49\textwidth}
    \includegraphics[trim=20mm 20mm 10mm 15mm,clip,width=\textwidth]{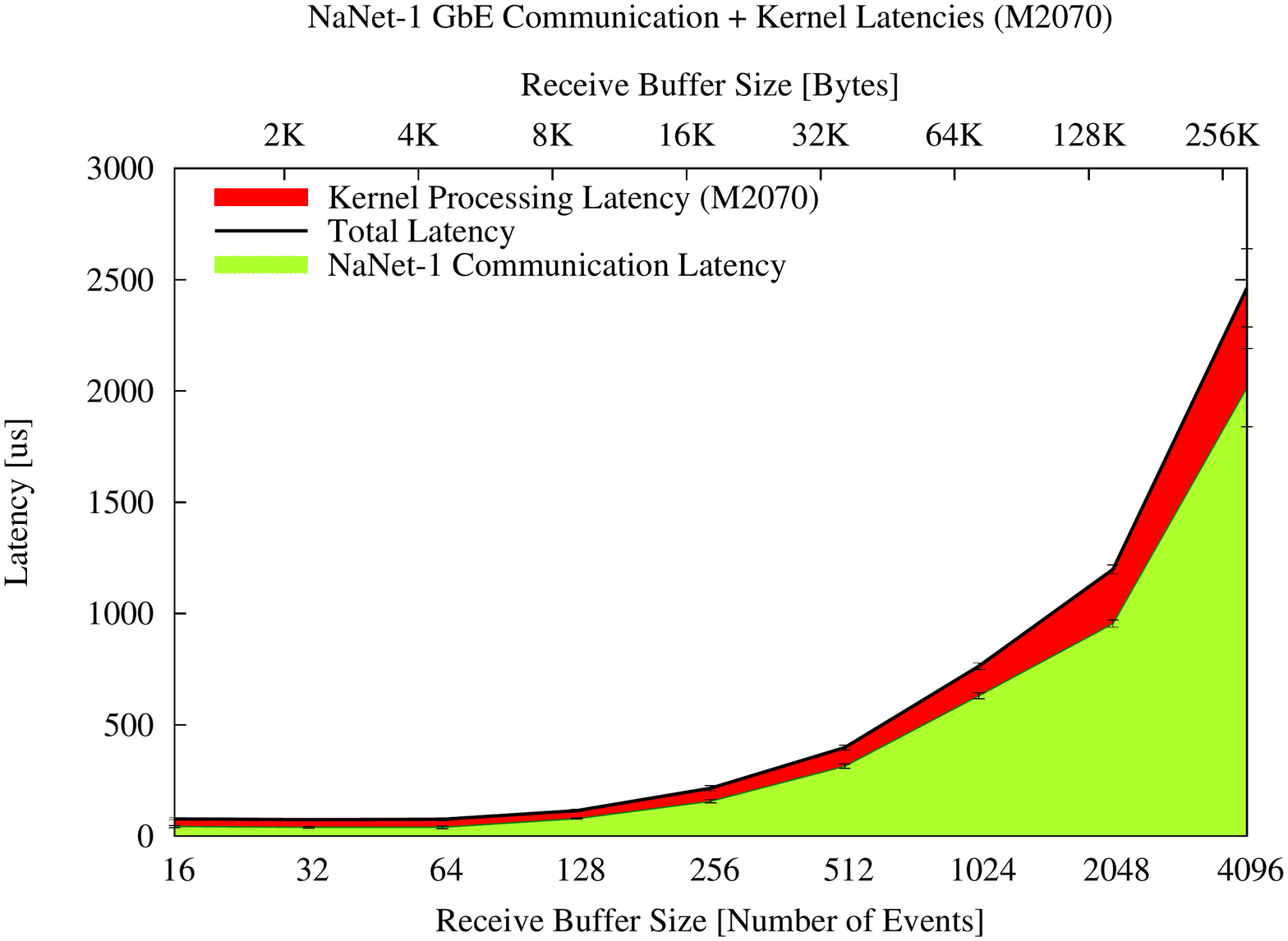}
    \caption {Latency of \nanetone \gbe data transfer and of ring
      reconstruction CUDA kernel processing.}
    \label{fig:latenza_nanet_fermi}
  \end{minipage}
  \vspace{-10pt}
\end{figure}
\begin{figure}[t]
  \begin{minipage}[t]{.49\textwidth}
    \includegraphics[trim=20mm 20mm 10mm 15mm,clip,width=\textwidth]{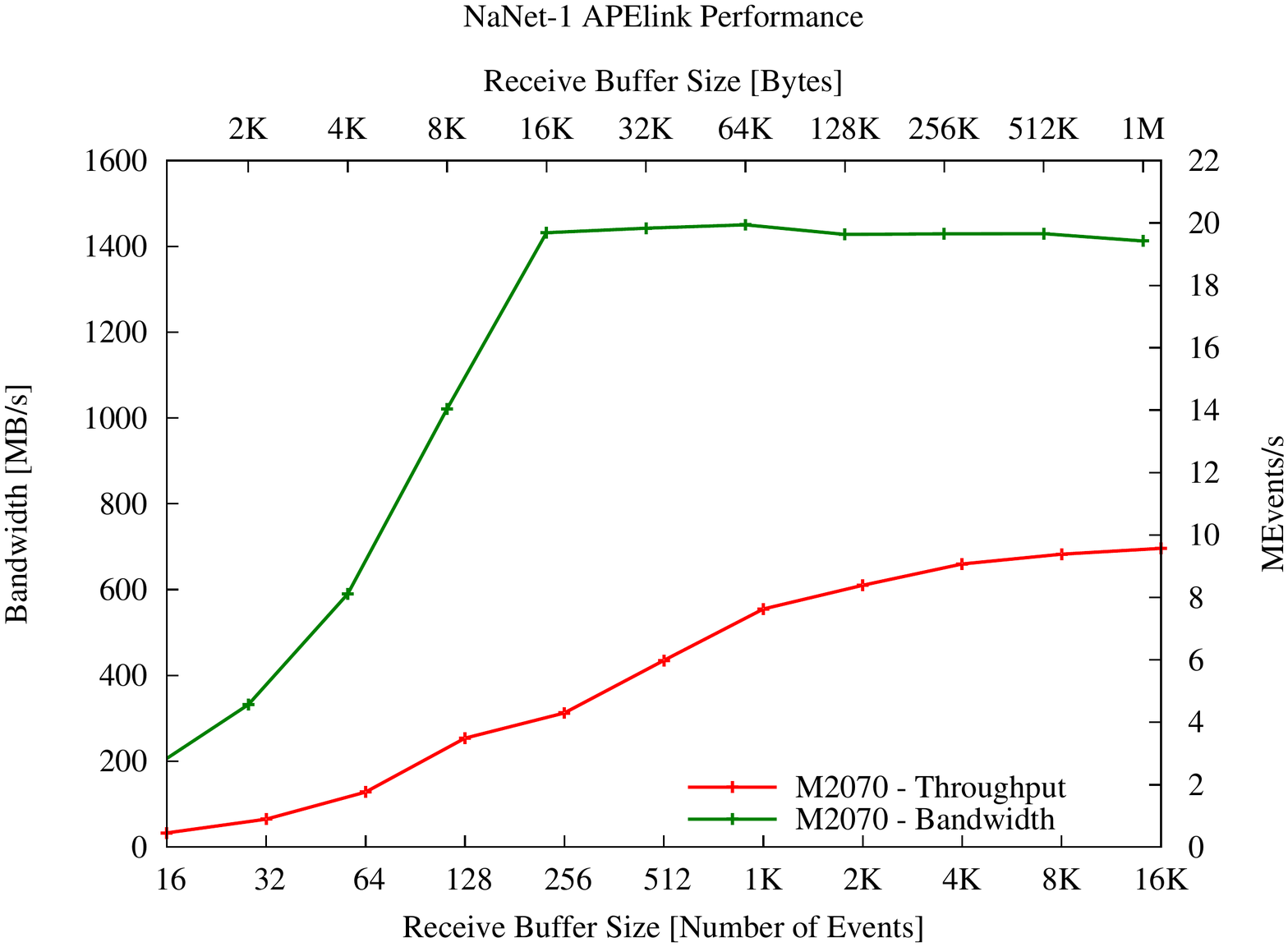}
    \caption {\nanetone \apelink bandwidth and \mbox{\apelink-fed}
      RICH L0TP throughput.}
    \label{fig:banda_throughput_apelink}
  \end{minipage}
  \quad
  \begin{minipage}[t]{.49\textwidth}
    \includegraphics[trim=20mm 20mm 10mm 15mm,clip,width=\textwidth]{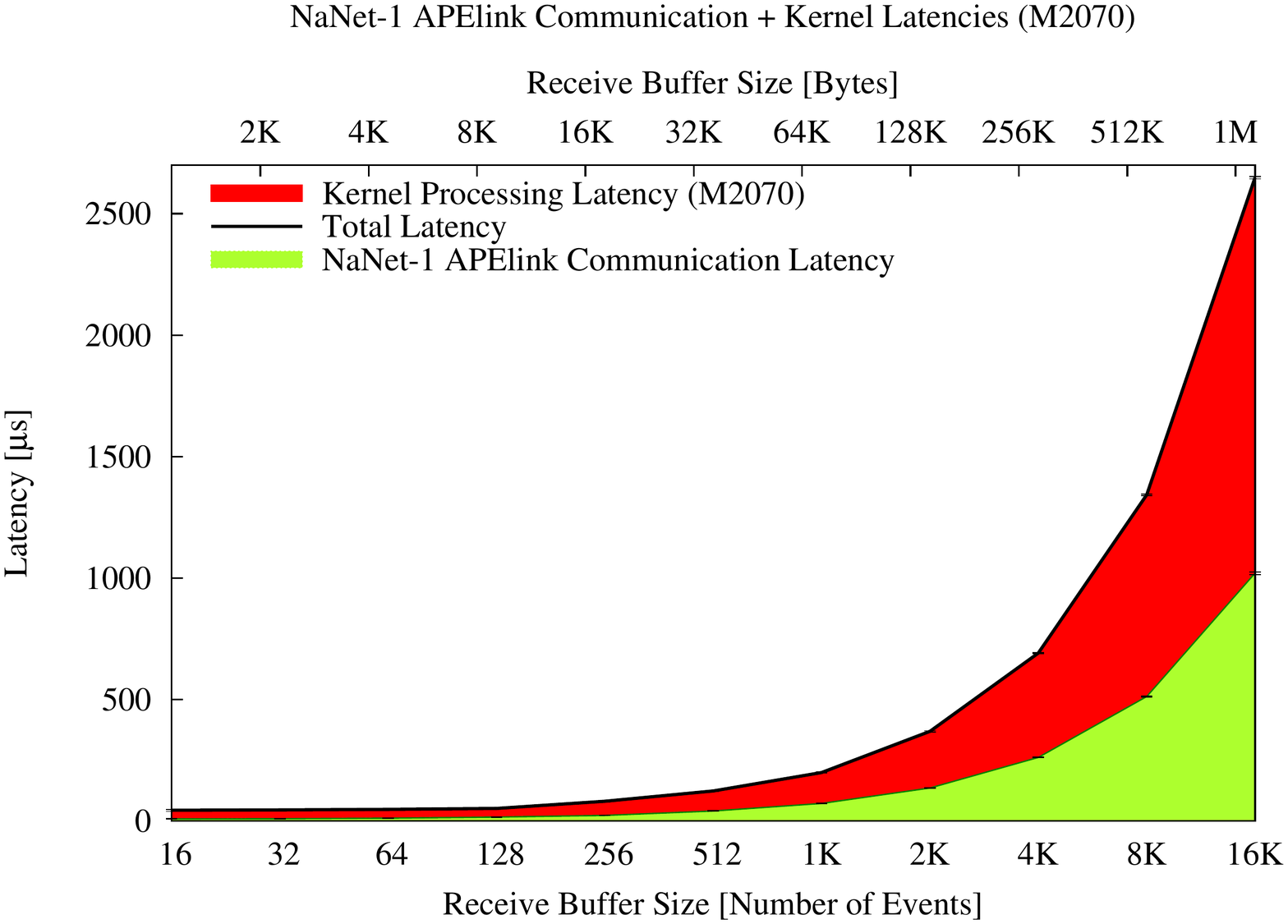}
    \caption {Latency of \nanetone \apelink data transfer and of ring
      reconstruction CUDA kernel processing.}
    \label{fig:latenza_apelink}
  \end{minipage}
  \vspace{-10pt}
\end{figure}

Benchmark results for \gbe link bandwidth, varying the size of GPU
memory receiving buffers, is shown in
figure~\ref{fig:banda_throughput_nanet}; it remains practically
constant in the region of interest for the reference application and
at maximum value for the link. 
In figure~\ref{fig:latenza_nanet_fermi} latencies for varying size
buffer transfers in GPU memory using the \gbe link are represented.
Besides the smooth behaviour increasing receive buffer sizes,
fluctuations are minimal, matching both constraints for \realtime and,
compatibly with link bandwidth, \lowlatency on data transfers; for a
more detailed performance analysis,
see~\cite{CHEP2013_arXiv:NANETshort}.

Bandwidth and latency performances for \nanetone \apelink channel are
in figure~\ref{fig:banda_throughput_apelink} and
figure~\ref{fig:latenza_apelink}.
Current implementation of \apelink is able to sustain a data flow
up to $\sim 20$~Gbps.
The \apelink bandwidth plateau in
figure~\ref{fig:banda_throughput_apelink} is due to the RX path
implementation of \nanetone.
\mbox{RDMA-related} tasks weigh on the \nios; for a $\sim 200$~MHz
clock, this means $\sim 1.6$~us more latency to each packet.

\section{The NA62 RICH Detector \mbox{GPU-Based} low level Trigger Case Study}
\label{sec:rich}
%
The NA62 experiment at CERN~\cite{Lamanna:2011zz} aims at measuring
the Branching Ratio (BR) of the \mbox{ultra-rare} decay of the charged
Kaon into a pion and a \mbox{$\nu\overline{\nu}$} pair.
Due to the very high precision of theoretical prediction on this BR, a
precise measurement at the level of 100 events would be a stringent
test of the Standard Model, also being this BR highly sensitive to any
new physics particle.
%

%
The $\sim 10$~MHz rate of particles reaching the detectors must be
reduced by a set of trigger levels down to a $\sim$~kHz rate,
manageable for data recording.
%
%
The first level (L0) is implemented in hardware (FPGAs) on the RO
boards and performs rough cuts on their output reducing $\sim10$ times
the data stream rate to cope with the $\leq1$~MHz event readout rate
for the design.
%
%
Events out from L0 are transferred for further reconstruction and
event building to upper level triggers (L1 and L2), implemented in
software on a farm of commodity PCs.
%
In the standard implementation, FPGAs on the L0 trigger RO boards
compute simple trigger primitives \mbox{on-the-fly} which are
\mbox{time-stamped} and sent to a central processor for matching and
trigger decision.
Thus, the maximum latency allowed for the synchronous L0 trigger is
related to the maximum data storage time available on the data
acquisition boards, up to 1~ms for NA62.
%
%
The Ring Imaging \v{C}erenkov detector (RICH) identifies pions and
muons in the momentum range 15~$GeV/c$ to 35~$GeV/c$, giving a $\mu$
suppression factor better than $10^{-2}$ with a good time resolution.
%
%
%
%
%

As a first example of GPU application in the NA62 trigger we studied
ring reconstruction in the RICH.
%
%
%
The RICH L0 trigger processor is a \lowlatency
synchronous level and the possibility to use the GPU must be verified.
In order to test feasibility and performances, as a starting point we
have implemented 5 algorithms for single ring finding in a sparse
matrix of 1000 points (centered on the PMs in the RICH spot) with 20
firing PMs (``hits'') on average.
Results of this study are available in~\cite{Collazuol:2012zz} and
show that GPU processing latency is stable and reproducible once data
are available in the device internal memory.

%
%
In order to fully characterize latency and throughput of the
\mbox{GPU-based} RICH L0 trigger processor (GRL0TP), we took into
account, besides \mbox{GPU-assisted} ring reconstruction, data
transfer needed to move primitives data from RO boards to GPU internal
memory through multiple (4$\div$6) \gbe links and the host \pcie bus.
The \nanetone NIC was integrated in the GRL0TP prototype, using the
``system loopback'' setup described in section~\ref{sec:perf}.
The host simulates the RO board by sending UDP packets containing
primitives data from the \gbe port of the hosting system to the \gbe
port the hosted \nanetone, which in turn streams data directly towards
a circular list of receive buffers in GPU memory that are sequentially
consumed by the CUDA kernel implementing the ring reconstruction
algorithm.
Communication and kernel processing tasks were serialized in order to
perform the measure; results are shown in
Fig.~\ref{fig:latenza_nanet_fermi}.
This represents a \mbox{worst-case} situation: during normal operation
given \nanetone RDMA capabilities, this serialization does not happen,
and kernel processing seamlessly overlaps with data transfer.
This is confirmed by throughput measurements in
figure~\ref{fig:banda_throughput_nanet}.
Combining the results, it is clear that the system remains within the
1~ms time budget with GPU receive buffer sizes in the
$128\div1024$~events range while keeping a $\sim1.7$~MEvents/s
throughput.
Although real system physical link and data protocol were used to show
the \realtime behaviour on \nanetone, we measured on a reduced
bandwidth single \gbe port system that could not match the
10~MEvents/s experiment requirement for the GRL0TP.

To demonstrate the suitability of \nanetone design for the
\mbox{full-fledged} RICH L0TP, we decided to perform equivalent
benchmarks using one of its \apelink ports instead of the \gbe one.
Results for throughput and latency of the \apelink-fed RICH L0TP are
shown in figure~\ref{fig:banda_throughput_apelink}
and~\ref{fig:latenza_apelink}: a single \nanetone \apelink data
channel between RICH RO and GRL0TP systems roughly matches
trigger throughput and latency requirements for receiving buffer size
in the $4\div$5~Kevents range.


\section{Conclusions and Future Work}
\label{sec:conclusions}

In this paper we presented the \nanet board family, a modular design of 
a \lowlatency NIC dedicated to \realtime GPU-based systems and supporting
a number of different physical links.
%
%

%
%
A performance analysis of the \nanetone board has been provided,
showing the \realtime features of its \gbe channel.

We demonstrated that using a single \nanetone \apelink channel to feed
the RICH L0 \mbox{GPU-based} trigger processor roughly fulfil latency and
throughput requirements of the system.
While adding a \apelink channel to the RO board is likely infeasible,
needing a major redesign, it encouragingly hints to the suitability of
the \nanetten as RICH L0 \mbox{GPU-based} trigger processor NIC.

\acknowledgments

This work was partially supported by the EU Framework Programme 7
EURETILE project, grant number 247846; R. Ammendola was supported by
MIUR (Italy) through the INFN SUMA project.
G. Lamanna, F. Pantaleo and M. Sozzi thank the GAP project, partially
supported by MIUR under grant RBFR12JF2Z ``Futuro in ricerca 2012''.


\end{document}